\documentclass[longauth]{aa}   % for long lists of affiliations 
\usepackage{color}
\definecolor{black}{rgb}{0,0,0}
\definecolor{red}{rgb}{1,0,0}
\definecolor{darkblue}{rgb}{0,0,0.7}
\definecolor{blue}{rgb}{0,0,1} 
\definecolor{green}{rgb}{0,0.5,0} 
\definecolor{orange}{rgb}{0.8,0.6,0} 
\definecolor{purple}{rgb}{1,0,1}

\usepackage{orcidlink}
\usepackage{longtable,array}
\usepackage{graphicx}
\usepackage{txfonts}
\usepackage[utf8]{inputenc} % to have accents
\begin{document} 

   \title{Constraints on Triton atmospheric evolution from occultations: 1989-2022}
    \subtitle{}

%----------------------------------------------------------------------------------------------------------
\author{B. Sicardy$^{\orcidlink{0000-0003-1995-0842}}$\inst{1}
%  Bruno.Sicardy@obspm.fr
% LESIA, Observatoire de Paris, Universit'e PSL, CNRS, Sorbonne Universit'e, 5 place Jules Janssen, 92190 Meudon, France
\and
A. Tej$^{\orcidlink{0000-0001-5917-5751}}$\inst{2}
%  anandmayee.tej@gmail.com
% Indian Institute of Space Science and Technology, Thiruvananthapuram, 695547, Kerala, India
\and
A. R. Gomes-Júnior$^{\orcidlink{0000-0002-3362-2127}}$\inst{3,4}
%  altairgomesjr@gmail.com
% Federal University of Uberlândia (UFU), Physics Institute, Av. João Naves de Ávila 2121, Uberlândia, MG 38408-100, Brazil
% Laboratório Interinstitucional de e-Astronomia - LIneA, Av. Pastor Martin Luther King Jr 126, Rio de Janeiro, RJ 20765-000, Brazil
\and
F. D. Romanov$^{\orcidlink{0000-0002-5268-7735}}$\inst{5}
%  filipp.romanov.27.04.1997@gmail.com
% American Association of Variable Star Observers (AAVSO), 185 Alewife Brook Parkway, Suite 410, Cambridge, MA 02138, USA
\and
T. Bertrand$^{\orcidlink{0000-0002-2302-9776}}$\inst{1}
%  tanguy.bertrand@lmd.ipsl.fr
% LESIA, Observatoire de Paris, Universit'e PSL, CNRS, Sorbonne Universit'e, 5 place Jules Janssen, 92190 Meudon, France
\and
N. M. Ashok\inst{6}
%  ashoknagarhalli@gmail.com
% Physical Research Laboratory, Ahmedabad, 380009, Gujarat, India
\and
E. Lellouch$^{\orcidlink{0000-0001-7168-1577}}$\inst{1}
%  emmanuel.lellouch@obspm.fr
% LESIA, Observatoire de Paris, Universit'e PSL, CNRS, Sorbonne Universit'e, 5 place Jules Janssen, 92190 Meudon, France
\and
B. E. Morgado$^{\orcidlink{0000-0003-0088-1808}}$\inst{7}
%  morgado.fis@gmail.com
% Universidade do Rio de Janeiro - Observat\'orio do Valongo, Ladeira do Pedro Antonio 43, Rio de Janeiro, RJ 20.080-090, Brazil
\and
M. Assafin$^{\orcidlink{0000-0002-8211-0777}}$\inst{7,4} 
%  massaf@ov.ufrj.br
% Universidade do Rio de Janeiro - Observat\'orio do Valongo, Ladeira do Pedro Antonio 43, Rio de Janeiro, RJ 20.080-090, Brazil
% Laboratório Interinstitucional de e-Astronomia - LIneA, Av. Pastor Martin Luther King Jr 126, Rio de Janeiro, RJ 20765-000, Brazil
\and
J. Desmars$^{\orcidlink{0000-0002-2193-8204}}$\inst{8,9}
%  josselin.desmars@obspm.fr
% Institut Polytechnique des Sciences Avancées IPSA, F-94200 Ivry-sur-Seine, France
% IMCCE, Observatoire de Paris, PSL Research University, CNRS, Sorbonne Université, Univ. Lille, F-75014 Paris, France
\and
J. I. B. Camargo$^{\orcidlink{0000-0002-1642-4065}}$\inst{10,4}
%  camargo@on.br
% Observatório Nacional/MCTIC, Rio de Janeiro, Brazil
% Laboratório Interinstitucional de e-Astronomia - LIneA, Av. Pastor Martin Luther King Jr 126, Rio de Janeiro, RJ 20765-000, Brazil
\and
Y. Kilic$^{\orcidlink{0000-0001-8641-0796}}$\inst{11}
%  yucelkilic1@gmail.com
% T\"UBITAK National Observatory, Akdeniz University Campus, Antalya 07058, Turkey
\and
J. L. Ortiz$^{\orcidlink{0000-0002-8690-2413}}$\inst{12}
%  ortiz@iaa.es
% Instituto de Astrof'{\i}sica de Andaluc'{\i}a (IAA-CSIC), Glorieta de la Astronom'{\i}a s/n. 18008-Granada, Spain
\and
R. Vieira-Martins$^{\orcidlink{0000-0003-1690-5704}}$\inst{10,4,7}
%  rvm@on.br
% Observatório Nacional/MCTIC, Rio de Janeiro, Brazil
% Laboratório Interinstitucional de e-Astronomia - LIneA, Av. Pastor Martin Luther King Jr 126, Rio de Janeiro, RJ 20765-000, Brazil
% Universidade do Rio de Janeiro - Observat\'orio do Valongo, Ladeira do Pedro Antonio 43, Rio de Janeiro, RJ 20.080-090, Brazil
\and
F. Braga-Ribas$^{\orcidlink{0000-0003-2311-2438}}$\inst{13,4}
%  felipebribas@gmail.com
% Federal University of Technology-Paraná (UTFPR/PPGFA), Curitiba, PR, Brazil
% Laboratório Interinstitucional de e-Astronomia - LIneA, Av. Pastor Martin Luther King Jr 126, Rio de Janeiro, RJ 20765-000, Brazil
\and
J. P. Ninan\inst{14}
%  indiajoe@gmail.com
% Tata Institute of Fundamental Research, Homi Bhabha Road, Colaba, Mumbai 400 005, India
\and
B. C. Bhatt\inst{15}
%  bcb@iiap.res.in
% Indian Institute of Astrophysics, II Block, Koramangala, Bangalore, 560034, India
\and
S. Pramod Kumar\inst{15}
%  pramodkpsree@gmail.com
% Indian Institute of Astrophysics, II Block, Koramangala, Bangalore, 560034, India
\and
V. Swain$^{\orcidlink{0000-0002-7942-8477}}$\inst{16}
%  vishwajeet.s@iitb.ac.in
% Department of Physics, Indian Institute of Technology Bombay, Powai, 400 076, India
\and
S. Sharma$^{\orcidlink{0000-0001-5731-3057}}$\inst{17}
%  saurabh@aries.res.in
% Aryabhatta Research Institute of Observational Sciences, Manora Peak, Nainital 263002, India
\and
A. Saha\inst{2}
%  anindya.s1130@gmail.com
% Indian Institute of Space Science and Technology, Thiruvananthapuram, 695547, Kerala, India
\and
D. K. Ojha\inst{14}
%  ojha@tifr.res.in
% Tata Institute of Fundamental Research, Homi Bhabha Road, Colaba, Mumbai 400 005, India
\and
G. Pawar$^{\orcidlink{0000-0003-3639-9052}}$\inst{18,19}
%  pawarganesh1076@gmail.com
% Akashmitra Mandal, Kalyan, 421301, Maharashtra, India
% Nicolaus Copernicus Astronomical Center, Polish Academy of Sciences, ul. Rabia'nska 8, 87-100 Toru'n, Poland
\and
S. Deshmukh\inst{18}
%  shishir.supernova@gmail.com
% Akashmitra Mandal, Kalyan, 421301, Maharashtra, India
\and
A. Deshpande\inst{18}
%  ameyaddeshpande@gmail.com
% Akashmitra Mandal, Kalyan, 421301, Maharashtra, India
\and
S. Ganesh\inst{6}
%  shashi@prl.res.in
% Physical Research Laboratory, Ahmedabad, 380009, Gujarat, India
\and
J. K. Jain\inst{6}
%  jinesh@prl.res.in
% Physical Research Laboratory, Ahmedabad, 380009, Gujarat, India
\and
S. K. Mathew\inst{20}
%  shibu@prl.res.in
% Udaipur Solar Observatory, Physical Research Laboratory, P.O. Box 198, Badi Road, 313001, Udaipur, India
\and
H. Kumar$^{\orcidlink{0000-0003-0871-4641}}$\inst{21,16}
%  harshkosli13@gmail.com
% Harvard College Observatory, Harvard University, 60 Garden St. Cambridge 02158 MA
% Department of Physics, Indian Institute of Technology Bombay, Powai, 400 076, India
\and
V.  Bhalerao$^{\orcidlink{0000-0002-6112-7609}}$\inst{16}
%  varunb@iitb.ac.in
% Department of Physics, Indian Institute of Technology Bombay, Powai, 400 076, India
\and
G. C. Anupama$^{\orcidlink{0000-0003-3533-7183}}$\inst{15}
%  gca@iiap.res.in
% Indian Institute of Astrophysics, II Block, Koramangala, Bangalore, 560034, India
\and
S. Barway$^{\orcidlink{0000-0002-3927-5402}}$\inst{15}
%  udhanshu.barway@iiap.res.in
% Indian Institute of Astrophysics, II Block, Koramangala, Bangalore, 560034, India
\and
A. Brandeker$^{\orcidlink{0000-0002-7201-7536}}$\inst{22}
%  alexis@astro.su.se
% Department of Astronomy, Stockholm University, AlbaNova University Center, 10691 Stockholm, Sweden
\and
H. G. Florén$^{\orcidlink{0000-0001-6994-9159}}$\inst{22}
%  floren@astro.su.se
% Department of Astronomy, Stockholm University, AlbaNova University Center, 10691 Stockholm, Sweden
\and
G. Olofsson\inst{22}
%  olofsson@astro.su.se
% Department of Astronomy, Stockholm University, AlbaNova University Center, 10691 Stockholm, Sweden
\and
G. Bruno$^{\orcidlink{0000-0002-3288-0802}}$\inst{23}
%  giovanni.bruno@inaf.it
% INAF, Osservatorio Astroﬁsico di Catania, Via S. Soﬁa 78, 95123 Catania, Italy
\and
Y. M. Mao$^{\orcidlink{0000-0003-3184-7733}}$\inst{24,25}
% maoyiming22@mails.ucas.ac.cn
% Key Laboratory for Research in Galaxies and Cosmology, Chinese Academy of Sciences, 80 Nandan Rd., Shanghai 200030, China
% School of Astronomy and Space Science, University of Chinese Academy of Sciences, 1 East Yanqi Lake Rd., Beijing 100049, P. R. China
\and
R. H. Ye$^{\orcidlink{0000-0002-2339-5581}}$\inst{24,25}
%  renhaoye@shao.ac.cn
% Key Laboratory for Research in Galaxies and Cosmology, Chinese Academy of Sciences, 80 Nandan Rd., Shanghai 200030, China
% School of Astronomy and Space Science, University of Chinese Academy of Sciences, 1 East Yanqi Lake Rd., Beijing 100049, P. R. China
\and
Q. Y. Zou$^{\orcidlink{0000-0001-5469-6443}}$\inst{24,25}
%  zouqy@bao.ac.cn
% Key Laboratory for Research in Galaxies and Cosmology, Chinese Academy of Sciences, 80 Nandan Rd., Shanghai 200030, China
% School of Astronomy and Space Science, University of Chinese Academy of Sciences, 1 East Yanqi Lake Rd., Beijing 100049, P. R. China
\and
Y. K. Sun$^{\orcidlink{0000-0002-3935-2666}}$\inst{24,25}
%  sunyk@bao.ac.cn
% Key Laboratory for Research in Galaxies and Cosmology, Chinese Academy of Sciences, 80 Nandan Rd., Shanghai 200030, China
% School of Astronomy and Space Science, University of Chinese Academy of Sciences, 1 East Yanqi Lake Rd., Beijing 100049, P. R. China
\and
Y. Y. Shen$^{\orcidlink{0000-0001-8636-2990}}$\inst{26,25}
%  shenyy@bao.ac.cn
% Key Lab of Space Astronomy and Technology, National Astronomical Observatories, Beijing 100101, P.R. China
% School of Astronomy and Space Science, University of Chinese Academy of Sciences, 1 East Yanqi Lake Rd., Beijing 100049, P. R. China
\and
J. Y. Zhao$^{\orcidlink{0000-0002-2770-3481}}$\inst{27}
%  runwuyu20090722@foxmail.com
% Shandong Astronomical Society, 180 West Wenhua Rd., Weihai, Shandong 264209, P. R. China
\and
D. N. Grishin\inst{28}
%  dmigrishin@gmail.com
% Vygonnaia street, house 2/1, flat 59, Ussuriysk, Primorsky Krai 692527, Russia
\and
L. V. Romanova$^{\orcidlink{0009-0006-8165-4613}}$\inst{29}
%  larisa.romanova.09october.1975@gmail.com
% Koshkina Street, house 19 k. 1, flat 210, Moscow, 115409, Russia
\and
F. Marchis$^{\orcidlink{0000-0001-7016-7277}}$\inst{30,31}
%  fmarchis@seti.org
% SETI Institute, 339 N Bernardo Ave Suite 200, Mountain View, CA 94043, USA
% Unistellar, 5 allée Marcel Leclerc, bâtiment B, Marseille, F-13008, France
\and
K. Fukui$^{\orcidlink{0000-0002-9297-5133}}$\inst{32}
% kfukui@dab.hi-ho.ne.jp
% Unistellar Citizen Scientist
\and
R. Kukita$^{\orcidlink{0000-0001-7029-644X}}$\inst{32}
% kukita.xr250@gmail.com 
% Unistellar Citizen Scientist
\and
G. Benedetti-Rossi$^{\orcidlink{0000-0002-4106-476X}}$\inst{33,4,1}
%  gugabrossi@gmail.com
% UNESP—S\~ao Paulo State University, Grupo de Dinâmica Orbital e Planetologia, CEP 12516-410, Guaratinguetá, SP, Brazil
% Laboratório Interinstitucional de e-Astronomia - Av. Pastor Martin Luther King Jr 126, Rio de Janeiro, RJ 20765-000, Brazil
% LESIA, Observatoire de Paris, Universit'e PSL, CNRS, Sorbonne Universit'e, 5 place Jules Janssen, 92190 Meudon, France
\and
P. Santos-Sanz$^{\orcidlink{0000-0002-1123-983X}}$\inst{12}
%  psantos@iaa.es
% Instituto de Astrof'{\i}sica de Andaluc'{\i}a (IAA-CSIC), Glorieta de la Astronom'{\i}a s/n. 18008-Granada, Spain
\and
N. Dhyani\inst{18}
%  ndhyani29@gmail.com
% Akashmitra Mandal, Kalyan, 421301, Maharashtra, India
\and
A. Gokhale\inst{18}
%  architgokhale28@gmail.com
% Akashmitra Mandal, Kalyan, 421301, Maharashtra, India
\and
A. Kate\inst{18}
%  cometkate@gmail.com
% Akashmitra Mandal, Kalyan, 421301, Maharashtra, India
}

\institute{LESIA, Observatoire de Paris, Universit\'e PSL, CNRS, Sorbonne Universit\'e, 5 place Jules Janssen, 92190 Meudon, France\\
\email{Bruno.Sicardy@obspm.fr}
\and
Indian Institute of Space Science and Technology, Thiruvananthapuram, 695547, Kerala, India
\and
Federal University of Uberlândia (UFU), Physics Institute, Av. João Naves de Ávila 2121, Uberlândia, MG 38408-100, Brazil
\and
Laborat\'orio Interinstitucional de e-Astronomia - LIneA, Av. Pastor Martin Luther King Jr 126, Rio de Janeiro, RJ 20765-000, Brazil
\and
American Association of Variable Star Observers (AAVSO), 185 Alewife Brook Parkway, Suite 410, Cambridge, MA 02138, USA
\and
Physical Research Laboratory, Ahmedabad, 380009, Gujarat, India
\and
Universidade do Rio de Janeiro - Observat\'orio do Valongo, Ladeira do Pedro Antonio 43, Rio de Janeiro, RJ 20.080-090, Brazil
\and
Institut Polytechnique des Sciences Avancées IPSA, F-94200 Ivry-sur-Seine, France
\and
IMCCE, Observatoire de Paris, PSL Research University, CNRS, Sorbonne Université, Univ. Lille, F-75014 Paris, France
\and
Observatório Nacional/MCTIC, Rio de Janeiro, Brazil
\and
T\"UBITAK National Observatory, Akdeniz University Campus, Antalya 07058, Turkey
\and
Instituto de Astrof\'{\i}sica de Andaluc\'{\i}a (IAA-CSIC), Glorieta de la Astronom\'{\i}a s/n. 18008-Granada, Spain
\and
Federal University of Technology-Paraná (UTFPR/PPGFA), Curitiba, PR, Brazil
\and
Tata Institute of Fundamental Research, Homi Bhabha Road, Colaba, Mumbai 400 005, India
\and
Indian Institute of Astrophysics, II Block, Koramangala, Bangalore, 560034, India
\and
Department of Physics, Indian Institute of Technology Bombay, Powai, 400 076, India
\and
Aryabhatta Research Institute of Observational Sciences, Manora Peak, Nainital 263002, India
\and
Akashmitra Mandal, Kalyan, 421301, Maharashtra, India
\and
Nicolaus Copernicus Astronomical Center, Polish Academy of Sciences, ul. Rabia\'nska 8, 87-100 Toru\'n, Poland
\and
Udaipur Solar Observatory, Physical Research Laboratory, P.O. Box 198, Badi Road, 313001, Udaipur, India
\and
Harvard College Observatory, Harvard University, 60 Garden St. Cambridge 02158 MA
\and
Department of Astronomy, Stockholm University, AlbaNova University Center, 10691 Stockholm, Sweden
\and
INAF, Osservatorio Astroﬁsico di Catania, Via S. Soﬁa 78, 95123 Catania, Italy
\and
Key Laboratory for Research in Galaxies and Cosmology, Chinese Academy of Sciences, 80 Nandan Rd., Shanghai 200030, China
\and
School of Astronomy and Space Science, University of Chinese Academy of Sciences, 1 East Yanqi Lake Rd., Beijing 100049, P. R. China
\and
Key Lab of Space Astronomy and Technology, National Astronomical Observatories, Beijing 100101, P.R. China
\and
Shandong Astronomical Society, 180 West Wenhua Rd., Weihai, Shandong 264209, P. R. China
\and
Vygonnaia street, house 2/1, flat 59, Ussuriysk, Primorsky Krai 692527, Russia
\and
Koshkina Street, house 19 k. 1, flat 210, Moscow, 115409, Russia
\and
SETI Institute, 339 N Bernardo Ave Suite 200, Mountain View, CA 94043, USA
\and
Unistellar, 5 allée Marcel Leclerc, bâtiment B, Marseille, F-13008, France
\and
Unistellar Citizen Scientist
\and
UNESP-S\~ao Paulo State University, Grupo de Dinâmica Orbital e Planetologia, CEP 12516-410, Guaratinguetá, SP, Brazil}
 
\date{Received mm:dd, yyyy; accepted mm:dd, yyyy}

%%%%%%%%%%%%%%%%%  ABSTRACT %%%%%%%%%%%%%%%%%%%%%%%%%%
%"The required sections are Aims, Method, and Results, even if you choose not to use the headings." - from A&A website. Not sure how to remove the headings though
% See abstract and keywords after the end of document: 
% issues with the front page that is too long -->
% The abstract and keywords are added "by hand" below

  \abstract
  % context heading (optional)
   {%
   Around the year 2000, Triton's south pole experienced an extreme summer solstice that occurs every 
   $\sim$650 years, when the subsolar latitude reached about 50$^\circ$S.
   Bracketing this epoch,
   a few occultations probed Triton's atmosphere in 1989, 1995, 1997, 2008 and 2017. 
   A recent ground-based stellar occultation observed on 6 October 2022 provides 
   a new measurement of Triton's atmospheric pressure which is presented here.
   }%
  % aims heading (mandatory)
   {%
   The goal is to constrain the Volatile Transport Models (VTMs) of Triton's atmosphere that is basically
   in vapor pressure equilibrium with the nitrogen ice at its surface. 
   }%
  % methods heading (mandatory)
   {%
   Fits to the occultation light curves yield Triton's atmospheric pressure at the reference radius 1400~km,
   from which the surface pressure is induced. 
   }%
  % results heading (mandatory)
   {%
   The fits provide a pressure $p_{1400}= 1.211 \pm 0.039$~$\mu$bar at radius1400 km (47-km altitude), 
   from which a surface pressure of
   $p_{\rm surf}= 14.54 \pm 0.47$~$\mu$bar is induced (1$\sigma$ error bars).
   To within error bars, this is identical to the pressure derived from the previous occultation of 5 October 2017, 
   $p_{1400}= 1.18 \pm 0.03$~$\mu$bar and $p_{\rm surf}= 14.1 \pm 0.4$~$\mu$bar, respectively.
   Based on recent models of Triton's volatile cycles, the overall evolution over the last 30 years of the surface 
   pressure is consistent with N$_2$ condensation taking place in the northern hemisphere.
   However, models typically predict a steady decrease in surface pressure for the period 2005-2060, 
   which is not confirmed by this observation. Complex surface-atmosphere interactions, such as ice albedo runaway 
   and formation of local N$_2$ frosts in the equatorial regions of Triton could  
   explain the relatively constant pressure between 2017 and 2022.
   }%
  % conclusions heading (optional)
   {}
   

   \maketitle

\section{Introduction}
\label{sec_intro}
Triton is the largest of Neptune's satellites. 
With the Saturnian satellite Titan, it is the only satellite known to possess a global atmosphere.
%It was first suspected from ground-based spectroscopic observation \citep{cru79}, and 
%Ground-based spectroscopic observation \citep{cru79} gave the first evidence of an atmosphere which was later
Triton's atmosphere was revealed during the NASA Voyager 2 (V2) flyby of Neptune's main satellite in August 1989.
The Radio Science Subsystem (RSS, \citealt{tyl89,gur95}) occultation provided a
surface pressure of $p_{\rm surf}=14 \pm 1$~$\mu$bar (1$\sigma$-level).
Combined with other V2 data and subsequent ground-based observations, it %appeared 
was seen that Triton's tenuous atmosphere is mainly composed of nitrogen,  
N$_2$, in vapor pressure equilibrium with the icy surface. 

Since 1989, a handful of Earth-based observations of stellar occultations monitored Triton's atmosphere. 
The 15 August 1995 \citep{olk97}, 18 July 1997 (\citealt{ell00} and \citealt{mar22}, MO22 hereafter),
and 4 November 1997 \citep{ell03} events indicated a significant increase of pressure relative to the RSS measurement.
No further constraints on Triton's atmospheric pressure could be achieved from the 21 May 2008 occultation, 
which had a grazing geometry (MO22). 
%An occultation observed on 21 May 2008 could not bring any further constraints on Triton's atmospheric pressure, 
% due to the grazing geometry of this event (MO22).

The 5 October 2017 ground-based occultation campaign provided the first dense coverage
of Triton's atmosphere, with 90 occultation chords scanning both hemispheres of the satellite (MO22).  
Combined with the good signal-to-noise ratio of some of the light curves, this event yielded strong constraints 
on the thermal, density and pressure profiles of this atmosphere between the altitude levels $\sim$8~km and $\sim$190~km
(from $\sim$9~$\mu$bar and a few nanobars, respectively).
In particular, the atmospheric pressure was found to be back to the RSS value of 1989.
Moreover, the detection and structure of a central flash revealed an essentially spherical atmosphere with an 
apparent oblateness of less than 0.0011 at the 8-km altitude level.

Triton has recently experienced a rare ``extreme southern solstice'' that occurs every $\sim$650 years; 
see details in \cite{ber22}.
The subsolar latitude on the satellite reached about 50$^\circ$S in 2000. 
% The monitoring of Triton's atmosphere from 1989 to 2022, bracketing this period, 
% aims at revealing seasonal effects occurring in Triton's atmosphere. 
Seasonal variations in surface pressure can then constrain Volatile Transport Models (VTMs) 
that account for volatile transport induced by insolation changes. 
These models assume that Triton’s atmosphere has a negligible radiative thermal influence 
on the energy balance of its surface.
They calculate the local insolation at the surface and the thermal infrared cooling, % (neglecting the atmosphere), 
the heat storage in regions covered by N$_2$ ice and conduction in the subsurface, and, 
in the presence of N$_2$ ice, the condensation-sublimation rates necessary 
to force the surface temperature to remain at the nitrogen frost point, 
which depends in turn on surface pressure. 
These processes are sufficient to estimate to first order the temporal evolution of 
the surface temperature, pressure and N$_2$ transport. 

The first numerical Triton's VTMs emerged in the late 1980s, motivated by the V2 Triton flyby
(\citealt{spe90}, \citealt{han92}, \citealt{spe92}, \citealt{bro94}).
%
%Owing to the similarities with Pluto's atmosphere (a N$_2$ atmosphere controlled by vapor pressure equilibrium 
%with surface ices), lessons were learned after the Pluto New Horizons flyby in 2015. 
%This allowed VTMs for Triton to be updated % as Triton's volatile cycles 
%in light of new observable constraints (MO22, \citealt{ber22}).
Triton's atmosphere has similarities with that of Pluto's (a N$_2$ atmosphere controlled by 
vapor pressure equilibrium with the surface ices). Hence, crucial insights were gained after 
the Pluto New Horizons flyby in 2015 which allowed the VTMs for Triton to be updated in light of 
new observable constraints (MO22, \citealt{ber22}). 

After the pressure increase noted in 1995 and 1997, 
the surface pressure in 2017 was found to be close to the V2 value (MO22). 
To first order, the volatile transport models have been able to reproduce this trend for a wide range of model parameters.
It stems from nitrogen ice sublimation peaking in the southern hemisphere in $\sim$2000-2005 
as the southern N$_2$ ice cap is under maximum insolation (summer solstice). 
However, \cite{ber22}'s models do not suggest a strong surge in surface pressure (see their Figs.~9, 10 and 21), 
as claimed from stellar occultations for the period 1995–1997 (e.g. \citealt{ell00}), 
but instead predict a moderate increase, % with much less amplitude, 
with a peak around 15–19~$\mu$bar for the surface pressure.
However, other processes not taken into account in these models (e.g. ice albedo feedback) 
may have increased the peak amplitude during this period.

Here, we present results obtained from a new stellar occultation event observed on 6 October 2022
from ground-based (China and India) and space (CHEOPS) facilities. 
These new results help constrain the evolution of Triton's atmosphere over the period 1989-2022.
This is a timely event, considering the rarity of Triton occultations, 
owing to the depleted stellar fields that Neptune's system is currently crossing as seen from Earth.

%%%%%%%%%%%%%%%%%%%%%%%%%%%%%%%%%%%%%%%%%%%%%%%%%%%%%%%%%%%%%%%%%%%%%%%%%%%%%%
%%%%%%%%%%%%%%%%%%%%%  Fig.  shadow path, geometry %%%%%%%%%%%%%%%%%%%%%%%%%%%
%%%%%%%%%%%%%%%%%%%%%%%%%%%%%%%%%%%%%%%%%%%%%%%%%%%%%%%%%%%%%%%%%%%%%%%%%%%%%%
\begin{figure}[h!]
\centerline{\includegraphics[width=\linewidth,trim=0 18cm 0 0, clip]{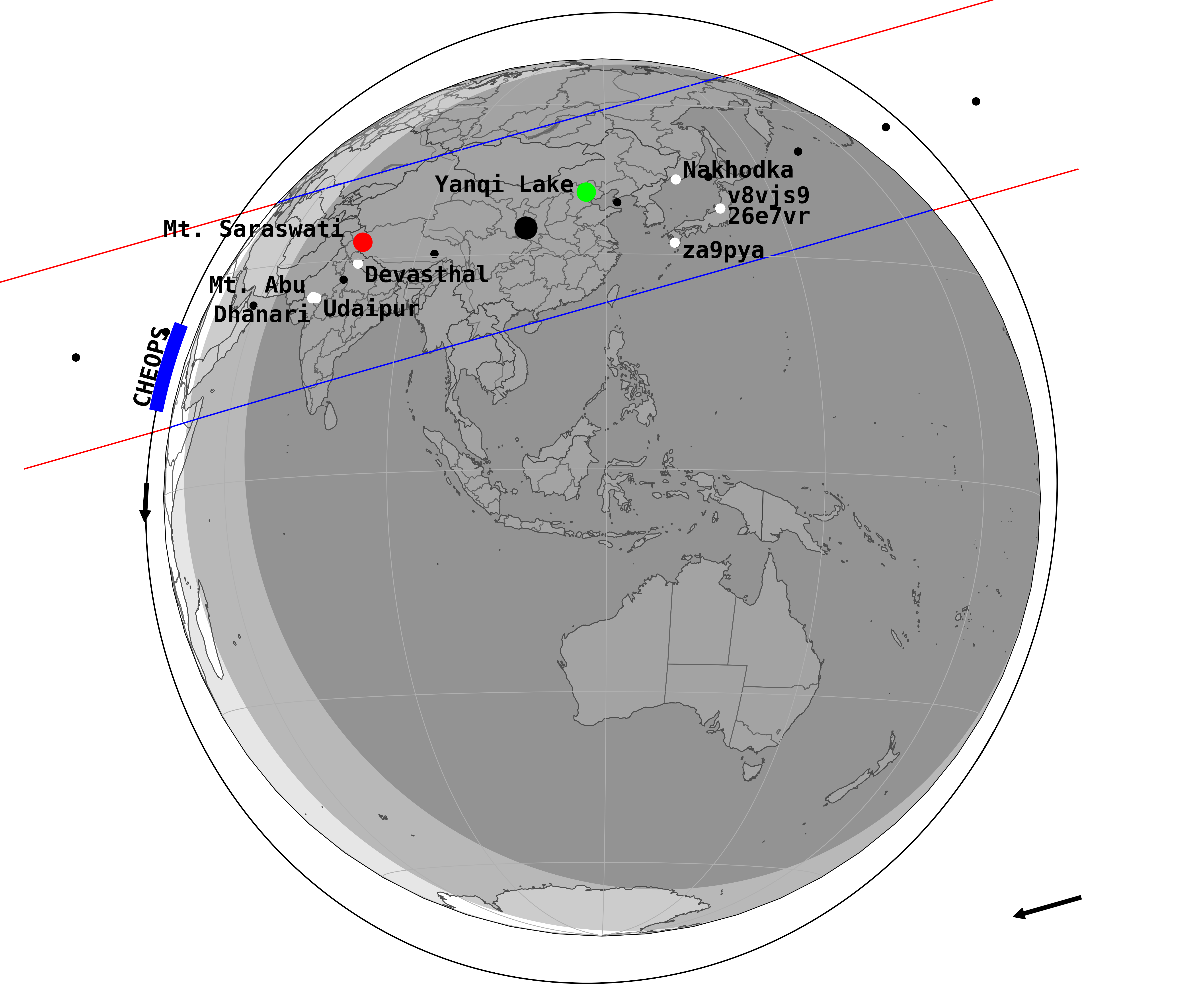}}
\centerline{\includegraphics[totalheight=60mm,trim=0 0 0 0]{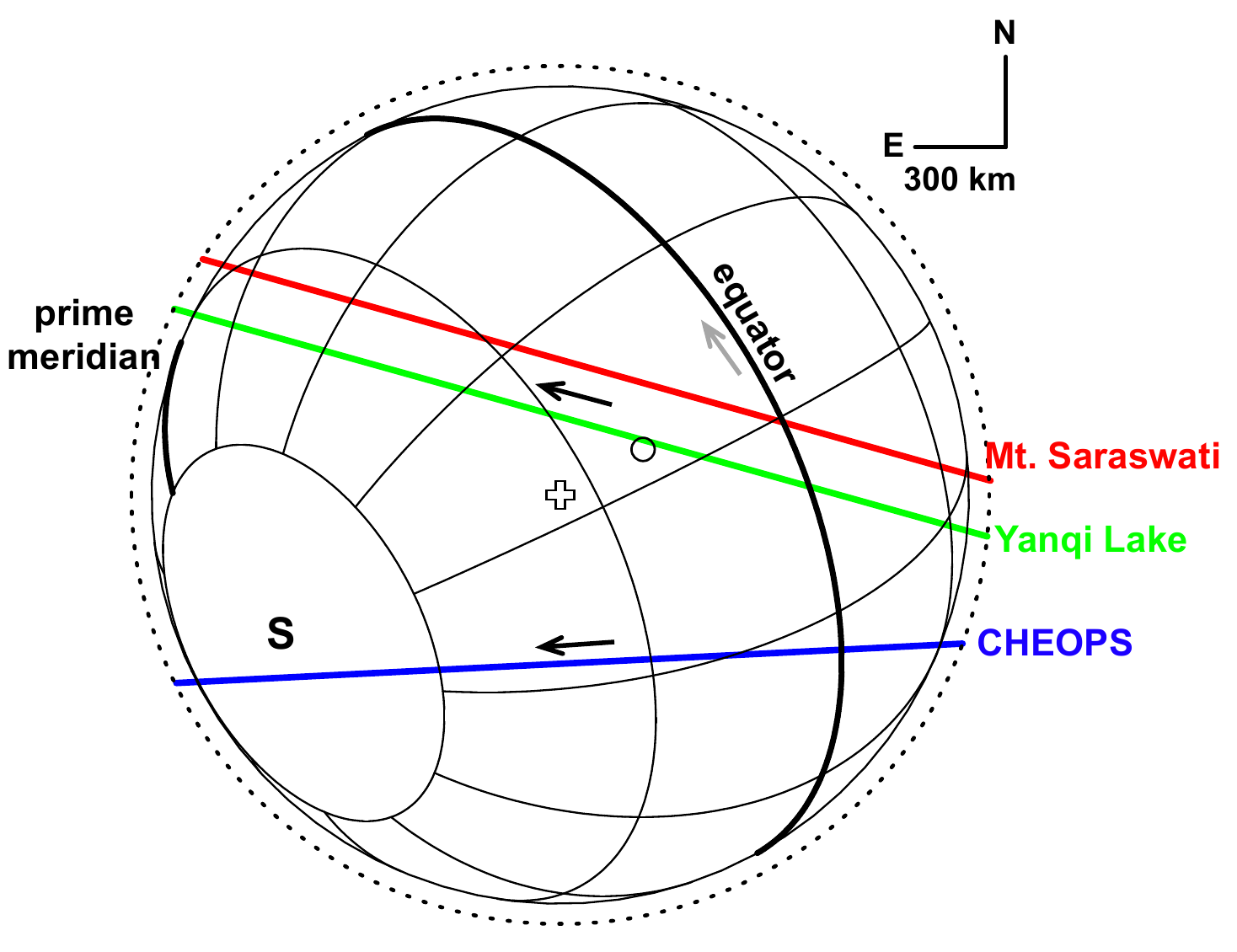}}
\caption{
\it Upper panel: \rm
The dark blue lines (continued in red outside the Earth) delimit the
predicted path of Triton's shadow on Earth on 6 October 2022.
The shadow motion is from right to left.
The shadow centers (black dots) are spaced by one minute, the larger black dot marking 
the geocentric closest approach near 14:49:46 UT (Table~\ref{tab_results}).
% while the thin light blue lines mark the shadow half-light level at radius 1420~km. % (1420~km) ?
%
Colors indicates stations where the occultation was successfully detected.
The blue arc shows the motion of the CHEOPS spacecraft during the event, while
the red and green dots are for Mt. Saraswati and Yanqi lake, respectively.
The white dots are for the stations that were clouded out, see Table~\ref{tab_observations}.
\it Lower panel: \rm
The reconstructed geometry of the occultation as seen in the sky plane.
The J2000 celestial north (N) and east (E) directions and the scale are displayed in the upper right corner.
The grey arrow near the equator shows the direction of rotation of the satellite. 
The (Neptune-facing) prime meridian is drawn as a thicker line compared to the other meridians, 
and the label S marks the south pole.
The dotted circle indicates the layer in Triton's atmosphere that causes the half-light level
($\sim$90-km altitude).
The colored lines are the trajectories of the star relative to Triton (or `occultation chords') 
as observed from various stations (see labels), 
with the black arrow indicating the direction of motion. 
% Blue: CHEOPS, green: Yanqi Lake, red: HCT and Growth telescopes at Mt. Saraswati.
The open circle is Triton's predicted center while the cross marks the actual center 
derived from the atmospheric fit.
The offset between the two mainly stems from a correction on Triton's ephemeris.
}
\label{fig_geometry}
\end{figure}
%%%%%%%%%%%%%%%%%%%%%%%%%%%%%%%%%%%%%%%%%%%%%%%%%%%%%%%%%%%%%%%%

\section{Observations}
\label{sec_occ_06oct22}

The Triton occultation campaign of 6 October 2022 was organized 
under the auspices of the Lucky Star project\footnote{https://lesia.obspm.fr/lucky-star/}. 
Comprehensive details regarding the event can be found in
a dedicated web page\footnote{https://lesia.obspm.fr/lucky-star/occ.php?p=109326}.
The compilation and management of the observational data are facilitated by Lucky Star's Occultation Portal 
website \footnote{https://occultation.tug.tubitak.gov.tr} \citep{kil22}.
The Gaia DR3 position at the epoch of occultation and the Triton's ephemeris from 
previous occultation events were used for the final prediction, see Fig.~\ref{fig_geometry}. 

The occultation was successfully observed from space by the CHaracterising ExOPlanet Satellite 
(CHEOPS\footnote{https://www.esa.int/Science\_Exploration/Space\_Science/Cheops}). 
% Earth-based observations were attempted from several sites in India, China and Russia. 
Two successful observations were obtained in India at Mt. Saraswati in the Himalayan region, 
while one light curve was successfully recorded at Yanqi Lake, China.
Attempts from Devasthal, Uttarakhand (India), Mt. Abu and Dhanari, Rajasthan (India), Udaipur (India), Nakhodka (Russia) 
and three sites in Japan were clouded out, 
see details in Appendix~\ref{ap_observations} and Table~\ref{tab_observations}.

%Observations were attempted from \red xxx \bla sites. The event was successfully recorded from four of them. One was achieved from space by the CHaracterising ExOPlanet Satellite 
%(CHEOPS\footnote{https://www.esa.int/Science\_Exploration/Space\_Science/Cheops}), two were obtained at Mt. Saraswati in the Indian Himalayan region, and one at Yanqi Lake, China. The successful campaigns are briefly discussed below and summarized in Table~\ref{tab_observations}.
%and below for details.

%%%%%%%%%%%%%%%%%%%%%%%%%%%%%%%%%%%%%%%%%%%%%%%%%%%%%%%%%%%%%%%%%%%%%%%%%%%%%
%%%%%%%%%%%%%%%%%%%%%% Table results %%%%%%%%%%%%%%%%%%%%%%%%%%%%%%%%%%%%%%%%
%%%%%%%%%%%%%%%%%%%%%%%%%%%%%%%%%%%%%%%%%%%%%%%%%%%%%%%%%%%%%%%%%%%%%%%%%%%%%
\begin{table*}[!t]
    \setlength{\tabcolsep}{2mm}
    \renewcommand{\arraystretch}{1}   % spacing between table lines
    \centering
    \caption{%
    Parameters of occulted star and Triton, and results of the atmospheric fits. Error bars are given at 1$\sigma$ level.
    \label{tab_results}
    }%
    \begin{tabular}{ll}
    \hline \hline

    \multicolumn{2}{c}{Occulted star} \\
    \hline
    Identification (Gaia DR3)     & 2639239368824994944 \\
    ICRS position at occultation epoch  & $\alpha= 23^{\rm h}36^{\rm m}52.45142^{\rm s} \pm 0.7$~mas, $\delta= -03^{\circ} 50^{\prime}09.7952^{\prime\prime} \pm 0.9$~mas \\

    \hline
    \multicolumn{2}{c}{Triton's parameters} \\
    \hline
    Mass, radius \tablefootmark{1}                       & $GM_{\rm T}= 1.4279 \times 10^{12}$  m$^3$ sec$^{-2}$, $R_{\rm T}= 1353$~km \\
    Geocentric distance                                  & $4.33409 \times 10^9$ km \\
    Pole position\tablefootmark{2}  (ICRS)              & $\alpha_{\rm p}$= 20h 13m 52.4s, $\delta_{\rm p}$=  20$^\circ$ 32' 38.2" \\
    Sub-solar latitude, sub-observer latitude, longitude & 34.2$^\circ$ S, 34.6$^\circ$ S, 115.0$^\circ$ E \\
    North pole position angle\tablefootmark{3}           & 298.2$^\circ$ \\
 
%    \hline
%    \multicolumn{2}{c}{Triton's atmosphere} \\
%    \hline
%    N$_2$ molecular mass    & $\mu= 4.652 \times 10^{-26}$ kg \\
%    N$_2$ molecular refractivity        & $K = 1.091 \times 10^{-23} + (6.282 \times 10^{-26}/\lambda_{\rm \mu m}^2)$ cm$^3$ molecule$^{-1}$ \\ 
%    Boltzmann constant      & $k= 1.380626 \times 10^{-23}$ J K$^{-1}$ \\

    \hline
    \multicolumn{2}{c}{Reconstructed geometry} \\    
    \hline
    Half-light layer in Triton's atmosphere, in  shadow plane       & 1442~km, 1420~km\\
% taken from positionv_CA_hct.i -->
%   Closest approach distance and time of HCT to shadow center      & $\rho_{\rm C/A,H}=$431$\pm$3~km,  $t_{\rm C/A,H}$=14:41:27.96$\pm$0.04 s UT \\
    Geocentric closest approach distance and time for shadow center  & $\rho_{\rm C/A,G}= 4067 \pm 3$~km, $t_{\rm C/A,G}=$ 14:39:46.26 $\pm$ 0.04 s UT \\

    \hline
    \multicolumn{2}{c}{Atmospheric results} \\
    \hline
    \multicolumn{2}{c}{This work} \\
    \hline
    Pressure at reference radius 1400 km, at surface\tablefootmark{4} & 
    $p_{1400}= 1.211 \pm 0.039$~$\mu$bar, $p_{\rm surf}= 14.54 \pm 0.47$~$\mu$bar \\
    \hline
    \multicolumn{2}{c}{Other works} \\
    \hline
    25 August 1989 (RSS occultation)\tablefootmark{5}           & $p_{\rm surf}= 14.1 \pm 1$~$\mu$bar \\
    25 August 1989 (RSS occultation)\tablefootmark{6}           & $p_{\rm 1400}= 1.0 \pm 0.2$~$\mu$bar \\
    14 August 1995 (ground-based occultation)\tablefootmark{7}  & $p_{1400}= 1.4 \pm 0.1$~$\mu$bar \\
    18 July 1997 (ground-based occultation)\tablefootmark{8}    & $p_{1400}= 2.23 \pm 0.28$~$\mu$bar \\
    18 July 1997 (ground-based occultation)\tablefootmark{6}    & $p_{1400}= 1.9^{+0.45}_{-0.30}$~$\mu$bar \\
    4 November 1997 (HST occultation)\tablefootmark{9}          & $p_{1400}= 1.759 \pm 0.016$~$\mu$bar \\ 
    21 May 2008 (ground-based occultation)\tablefootmark{6}     & $p_{1400}= 1.15^{+1.03}_{-0.37}$~$\mu$bar \\
    5 October 2017 (ground-based occultation)\tablefootmark{6}  & $p_{1400}= 1.18 \pm 0.03$~$\mu$bar \\
    \hline
    \end{tabular}
\tablefoot{
\tablefoottext{1}{\cite{mck95}, where $G$ is the constant of gravitation.}
\tablefoottext{2}{On 6 October 2022, using \cite{arc18}}
\tablefoottext{3}{Position angle of Triton's north pole projected in the sky plane. 
Counted positively from celestial north to celestial east.}
\tablefoottext{4}{Assuming $p_{\rm surf}/p_{\rm 1400}$=12.01, see text.}
\tablefoottext{5}{\cite{gur95}.}
\tablefoottext{6}{MO22.}
\tablefoottext{7}{\cite{olk97}.}
\tablefoottext{8}{\cite{ell00}.}
\tablefoottext{9}{HST: Hubble Space Telescope, \cite{ell03}.}
}
\end{table*}
%%%%%%%%%%%%%%%%%%%%%%%%%%%%%%%%%%%%%%%%%%%%%%%%%%%%%%%%%%%%%%%

%%%%%%%%%%%%%%%%%%%%%%%%%%%%%%%%%%%%%%%%%%%%%%%%%%%%%%%%%%%%%%%%%%%%%%%%
%%%%%%%%%%%%%%%%%%%%%  Fig.  best fit to light curves %%%%%%%%%%%%%%%%%%
%%%%%%%%%%%%%%%%%%%%%%%%%%%%%%%%%%%%%%%%%%%%%%%%%%%%%%%%%%%%%%%%%%%%%%%%
\begin{figure}[h!]
\centerline{\includegraphics[totalheight=125mm]{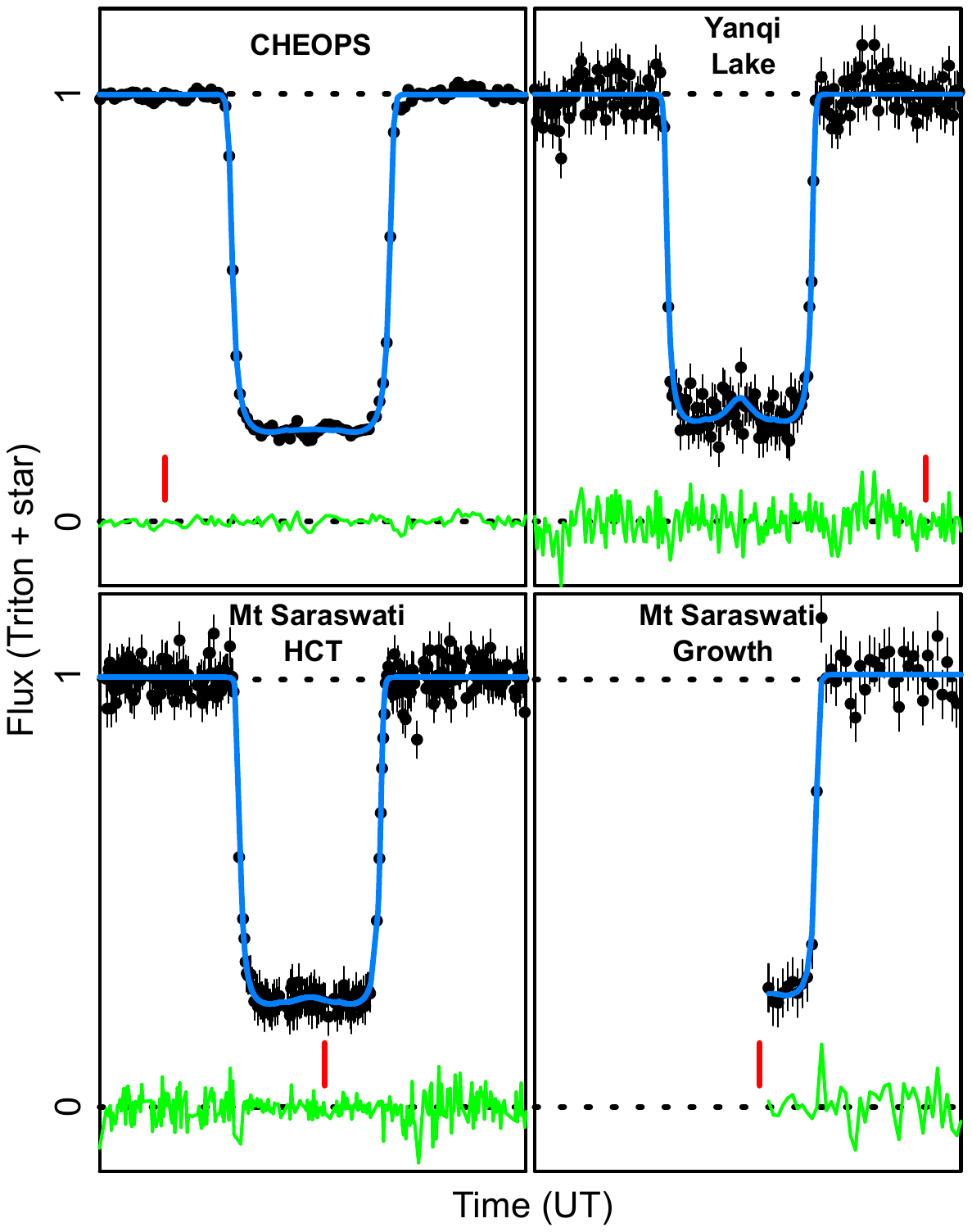}}
\centerline{\includegraphics[totalheight=55mm,trim=50 10 50 10]{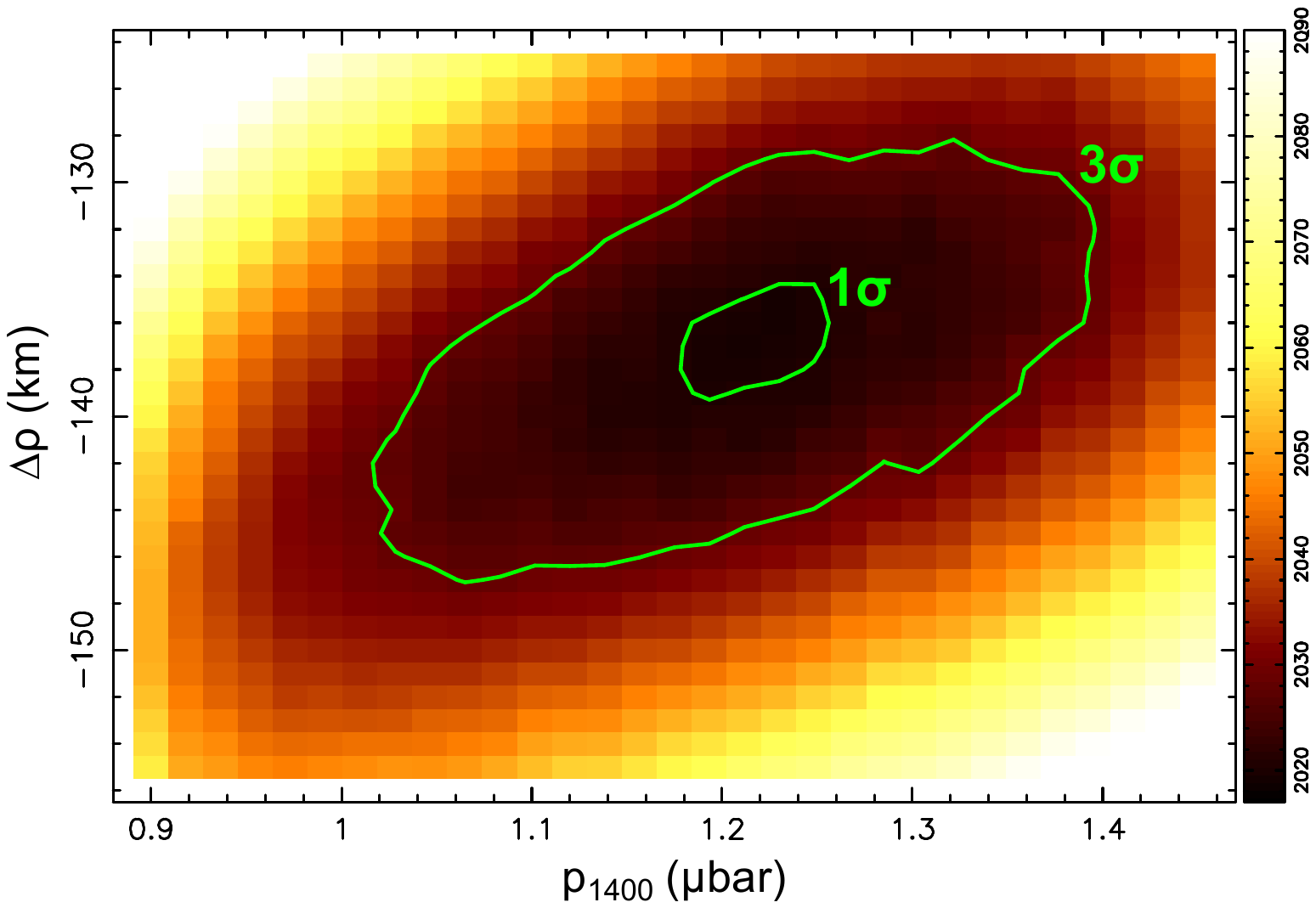}}
\caption{
\it Upper panel: \rm
the best fit to the data is shown as blue lines, using a Triton surface pressure of
$p_{\rm surf}= 14.54$~$\mu$bar and other parameters provided in Table~\ref{tab_results}.
The residuals (observations minus model) are shown in green.
The lower and upper horizontal dotted lines mark the zero flux and the average flux 
of Triton plus the star, respectively.
Each light curve is normalized to the total flux of Triton + the star and 
is plotted over a time interval of 6 minutes, the red tick mark indicating 14:41:40~UT.
The HCT light curve has been binned over five data points (1.05-second time resolution) 
for a better comparison with the other light curves.
We note that a faint central flash is present in the Yanqi Lake light curve.
%
%\blu
    The 1$\sigma$ error bars are obtained from the PRAIA software 
that accounts for the instrumental and photon noises,
knowing that the error bars for the CHEOPS data points are dominated by the photon noise (see Appendix~\ref{ap_observations}).
They are not visible here as they are smaller than the plotted dots.
%\bla
%
\it Lower panel: \rm
The $\chi^2$ map of the simultaneous fits, with the 1$\sigma$ and 3$\sigma$ level curves
shown in green.
}
\label{fig_best_fit}
\end{figure}
%%%%%%%%%%%%%%%%%%%%%%%%%%%%%%%%%%%%%%%%%%%%%%%%%%%%%%%%%%%%%%%%

\section{Data Analysis}

Our analysis used tools developed in the 
``Platform for Reduction of Astronomical Images Automatically"
(PRAIA, \citealt{ass23a,ass23b})\footnote{https://ov.ufrj.br/en/PRAIA/} and the packages of the
``Stellar Occultation Reduction and Analysis" (SORA, \citealt{gom22})
\footnote{https://sora.readthedocs.io/}

The resulting occultation light curves have been fitted using the ray-tracing code 
initially developed for Pluto's atmosphere by \cite{dia15}, see also \cite{mez19} and \cite{sic21}.
This code has been adapted to Triton's atmosphere, using the parameters of MO22 (their Table~2)
and in Table~\ref{tab_results} of this paper.
Triton's atmosphere is assumed to be transparent and composed of pure N$_2$,
noting that the CH$_4$ abundance of $10^{-4}$ is negligible for our purposes.
Further, it is assumed to be spherical, as supported by the shape of the central flash observed
during the 5 October 2017 event (MO22).
%, see the Introduction.
 
Triton's atmospheric thermal profile $T(r)$ -- where $r$ is the distance to Triton's center --
is the same as the one used by MO22, see their Figs.~10, 11 and B.1 and their Table~B.1.
It starts from the surface with a strong positive thermal gradient of 5 K~km$^{-1}$. 
This gradient decreases rapidly and the temperature reaches 
a local maximum value of about 50~K at $r=1363$~km (10~km altitude). 
This is followed by a mesosphere with a mild negative gradient of about -0.2 K~km$^{-1}$ 
centered around $r=1375$~km (23 km altitude). This mesosphere finally connects itself with a thermosphere 
that has a positive gradient of about 0.1 K~km$^{-1}$ above the 50-km altitude level.

A $\chi^2$-minimization procedure is applied by simultaneously fitting all the four light curves.
The $M=7$ free parameters of the fit are 
the pressure $p_{1400}$ at the reference radius $r=1400$~km, 
the offset $(f_{\rm c},g_{\rm c})$ to apply to Triton's ephemeris
along the celestial east and north directions, respectively, and
the contributions of Triton's flux to the total Triton + star flux in the four observed light curves.
These contributions are different from station to station because the instruments used different 
filters (Table~\ref{tab_observations}), inducing variations in the magnitudes of both Triton and the star.
The offset $(f_{\rm c},g_{\rm c})$ is obtained by determining 
the shift in time $\Delta t$ to apply to all the light curves and 
the cross-track offset to Triton's ephemeris, $\Delta \rho$,
that minimize $\chi^2$.

The fit uses a total $N=2398$ data points, providing a $\chi^2$ value per degree of freedom 
$\chi^2_{\rm dof}= \chi^2_{\rm min}/(N-M)$,
where $\chi^2_{\rm min}$ is the minimum value of $\chi^2$ obtained in the fitting procedure.
The best fit to the data (Fig.~\ref{fig_best_fit}) has a value $\chi^2_{\rm dof}= 0.84$, 
indicating a satisfactory modeling of the data.
The figure~\ref{fig_pres_evolution} shows the map of $\chi^2$ vs. 
the pressure $p_{1400}$ at radius 1400~km and the cross-track offset $\Delta \rho$.
As summarized in Table~\ref{tab_results}, the fit provides a best-fitting pressure of
% at the 1$\sigma$ and 3$\sigma$ levels, respectively.
$p_{1400}= 1.211 \pm 0.039$~$\mu{\rm bar}$ (1$\sigma$ level).
Our adopted temperature profile implies that $p_{\rm surf}/p_{1400}= 12.01$, yielding
%$p_{\rm surf}= 14.54 (\pm 0.53)(\pm 2.57)$~$\mu{\rm bar}$.
$p_{\rm surf}= 14.54 \pm 0.47$~$\mu{\rm bar}$.

\section{Pressure Evolution}
\label{sec_pres_evolution}

%%%%%%%%%%%%%%%%%%%%%%%%%%%%%%%%%%%%%%%%%%%%%%%%%%%%%%%%%%%%%%%%%%%%%%%%
%%%%%%%%%%%%%%%%%%%%%  Fig. pressure evolution %%%%%%%%%%%%%%%%%%%%%%%%%
%%%%%%%%%%%%%%%%%%%%%%%%%%%%%%%%%%%%%%%%%%%%%%%%%%%%%%%%%%%%%%%%%%%%%%%%
\begin{figure*}[h!]
\centerline{%
\includegraphics[totalheight=90mm,trim=0 0 0 0]{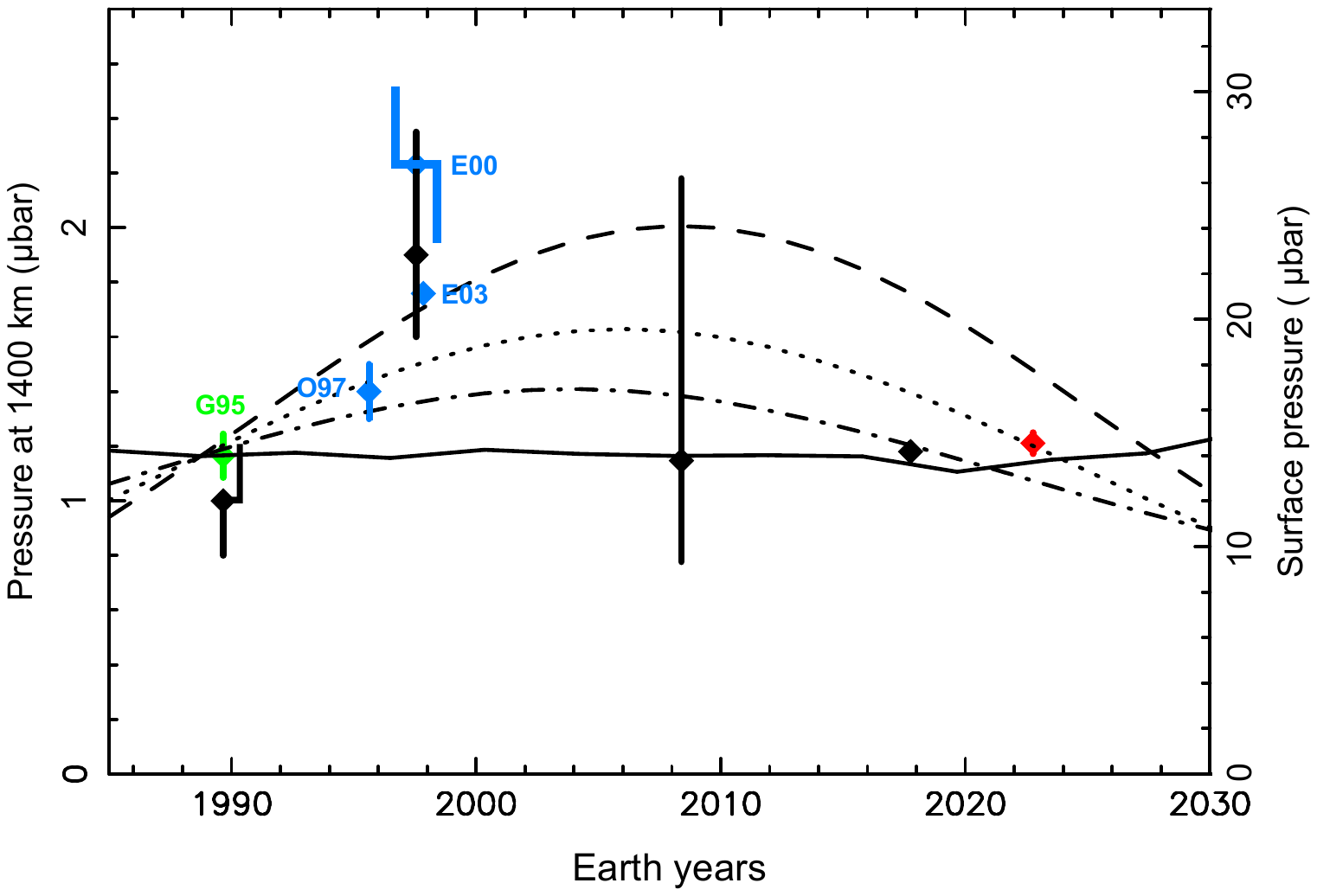}
}%
\caption{
Evolution of Triton's atmospheric pressure from occultation measurements taken between 1989 and 2022;
see dates and values in Table~\ref{tab_results}. 
The red point is the result of the 6 October 2022 occultation (this work).
It has been obtained using the same method as adopted by MO22 (black points).
The green point (G95) is from the V2 RSS occultation of 25 August 1989 \citep{gur95}.
The blue points are respectively:
O97, ground-based stellar occultation, 14 August 1995 \citep{olk97};
E03, ground-based stellar occultation, 18 July 1997 \citep{ell00};
E00, HST-based stellar occultation, 4 November 1997 \citep{ell03}.
It should be noted that MO22 analyzed the RSS data of August 1989 and the occultation data of July 1997
independently of \cite{gur95} and \citep{ell00}, respectively.
The vertical axes show the pressures at the reference radius 1400~km (left) and at the surface (right),
assuming $p_{\rm surf}/p_{1400}= 12.01$, see text.
The lines are examples of VTM simulations by \cite{ber22}.
% Dashed-dotted line: \blu \tt ps\_capC\_-90\_-30\_45\_90\_TI2000\rm\bla;
% dotted line: \blu \tt ps\_capC\_-90\_-30\_90\_90\_TI2000\rm\bla;
% dashed line: \blu \tt ps\_capC\_-90\_-30\_60\_90\_TI500\rm\bla;
% solid line: \blu \tt ps\_TI1000\_R1500\_A09\rm\bla.
%
Dashed-dotted line: southern cap within 90$^\circ$S-30$^\circ$S, northern cap within 45$^\circ$N-90$^\circ$N;
Dotted line: southern cap within 90$^\circ$S-30$^\circ$S, northern cap within 60$^\circ$N-90$^\circ$N;
Dashed line: southern cap within 90$^\circ$S-30$^\circ$S, no northern cap (volatile-free northern hemisphere);
solid line: a simulation where the N$_2$ ice freely evolves over millions of years so that permanent and seasonal polar caps as well as local frosts deposits can form self-consistently. In this simulation, formation of thin seasonal N$_2$ frosts is predicted in the current southern summer in the equatorial regions.
None of the simulations satisfactorily fits the data.
The dashed line accounts for the pressure increase of 1995-1997 but overestimates the pressures measured in 2017 and 2022.
The dotted line marginally explains the increase of 1995-1997 and satisfactorily fit the 2022 measurement, 
but fails to explain the 2017 point.
The dashed-dotted line does not fit the increase of 1995-1997, while going through the 2017 point,
but fails to explain the 2022 point.
The solid line accounts for a basically constant pressure between 1989, 2017 and 2022, 
as well as the slight increases between 2017 and 2022. 
However, it is only marginally consistent with the 1995 result, and 
does not account for the pressure surge of 1997.
}
\label{fig_pres_evolution}
\end{figure*}
%%%%%%%%%%%%%%%%%%%%%%%%%%%%%%%%%%%%%%%%%%%%%%%%%%%%%%%%%%%%%%%%

Figure~\ref{fig_pres_evolution} shows the measured values of Triton's atmospheric pressure 
over the period 1989-2022, together with various VTMs outputs of \cite{ber22}.
We consider here the simulations of these authors performed with unlimited and fixed N$_2$ ice reservoirs (i.e. no short-term frosts are involved) in the southern and northern hemispheres (see their Figs.~9-10), with some cases shown as dotted, dashed-dotted and dashed lines in Fig.~\ref{fig_pres_evolution}.

These simulations inevitably predict a steady decrease in pressure between 2005 and 2022,
with a pressure in 2022 that is at least 5\% lower than that in 2017 
(and 5\% in the case of the dash-dotted line in Fig.~\ref{fig_pres_evolution}),
followed by a steady decrease that is predicted to last for many decades.
This is because, as the subsolar point moves from the southern latitudes toward the equator in 2022,
nitrogen sublimation becomes less intense over the southern reservoir of N$_2$ ice, while condensation 
in the northern hemisphere and in the equatorial regions dominate sublimation at the global scale. 
%
% This trend is also obtained in their simulations performed with seasonal frost 
%\blu
This trend is also obtained in their simulations performed with limited N$_2$ ice reservoirs 
and with seasonal frosts, 
%\bla
that best match the observational constraints (see their Fig. 21), including among others 
the surface pressure in 1989 and 2017 and the latitudinal extent of the ice as suggested from V2 images. 

None of the simulations mentioned above can explain both the pressure surge observed in the years 1990’s and the return to the pressure of 1989 in 2017 and 2022 (Fig.~\ref{fig_pres_evolution}).

Some simulations of \cite{ber22} do obtain a relatively constant surface pressure between 1989, 2017 and 2022 that is consistent with our present result (Fig.~\ref{fig_pres_evolution}, solid line).
In these simulations, the N$_2$ condensation is more intense in the northern hemisphere during this period because the seasonal northern cap is strongly extended equatorward, in the form of tens-of-cm-thick frosts, due to a cold bedrock (typically, the bedrock surface bolometric albedo in these simulations is $\ge 0.9$). 
This balances the sublimation in the southern hemisphere and cause the pressure 
to remain relatively constant between 1989 and 2022. 
However, these simulations 
(1) include the presence of extended N$_2$ ice frosts % 0$^\circ$-45$^\circ$N 
to low northern latitudes in 1989, which seems inconsistent with the V2 images, 
unless the blue fringe observed in the equatorial regions correspond to a cm-thick N$_2$ frost 
(see discussions about the blue fringe in \citealt{ber22}, their section 8.4), and in fact 
(2) they do not include a peak in pressure at all during the 1995-1997 period.

Nevertheless, this surge should be considered with caution.
The black points in Fig.~\ref{fig_pres_evolution}, taken from MO22, 
are all obtained using the same temperature profile $T(r)$ and the same ray-tracing code.
In contrast, the blue points have been obtained with methods varying from each other and from our approach.
For instance, using the same data set stemming from the 18 July 1997 occultation, \cite{ell00} and MO22
derive $p_{1400} = 2.23 \pm 0.28$~$\mu$bar and $p_{1400} = 1.90^{+0.45}_{-0.30}$~$\mu$bar, respectively.
Although these two values agree to within their respective error bars, 
they show that a consistent comparison between results obtained by various teams remains problematic.
In that context, if we consider only the values 
$p_{1400} = 1.0 \pm 0.2$~$\mu$bar and
$p_{1400} = 1.90^{+0.45}_{-0.30}$~$\mu$bar 
derived by MO22 for the 25 August 1989 and 18 July 1997 measurements, respectively,
then the surge of pressure between 1989 and 1997 reaches 
%\blu 
a moderate 2.5$\sigma$ level. 
%\bla
% the 2.5$\sigma$ level only, which is marginally significant.

%\blu
This said, the formation of local N$_2$ frosts in the equatorial region during the period 1980-2020 
-- which is very sensitive to surface properties or albedo feedback -- 
may not be simulated in the VTM with sufficient details and could alter the pressure cycle.
Another possibility is that albedos and ice composition feedback may have a significant role on Triton 
as geysers or haze particles could deposit dark material (resp. bright ice grain) on top of the ice, 
and further darken (resp. brighten) the ice, and thus impact the N$_2$ sublimation-condensation rates. 
These processes, not taken into account in the VTM, have been shown to be efficient runaway forcing mechanisms 
on Pluto \citep{ear18,ber20}. They could be responsible for significant sub-seasonal
atmospheric pressure variations occurring in Pluto's atmosphere on timescales ranging from a few to tens 
of Earth years, as suggested by stellar occultations, see  \cite{ari20} and \citep{yua23}.
The latter authors note that short-term changes in Pluto’s surface ices have been reported
\cite{grun14,lell22,holl22}.
This could reconcile with the observations presented here, 
in particular with the 1995-1997 surge of pressure of Triton's atmosphere.
However, all the points evoked above remain speculative and should be explored in details with the models.
%\bla

\section{Conclusions}
\label{sec_conclusion}

The main constraint obtained here is that the Triton's atmospheric pressure is 
essentially the same in 1989, 2017 and 2022, whatever occurred in between.
More specifically, the pressure changed very little between 2017 and 2022 ,
with a non-significant increase of 
$0.031 \pm 0.049$~$\mu$bar (resp. $0.38 \pm 0.59$~$\mu$bar)
between the two dates at 1400~km (resp. the surface).

It remains difficult to clearly assess the seasonal trend of the N$_2$ cycle on Triton, 
because the simulations from \cite{ber22} do not reconcile all the available observations. 
i.e all pressure measurements over the 1989-2022 period and the visual appearance 
of Triton's surface in Voyager~2's images.
Another issue is that our 2022 measurement is close in time to the 2017 measurement, 
compared to a seasonal timescale on Triton ($\sim$40 years). 
Complex processes, not taken into account in the VTMs, could occur and 
perturb the N$_2$ cycle over a short timescale of 5 years, 
thus temporarily masking the seasonal trend,
%\blu 
see the discussion at the end of the previous Section. 
%\bla
%
% \red
% [respond to Emmanuel's comment: ``This would accentuate the pressure decrease from 2017 to 2022,
% contrarily to what is observed."] ->
% [Not necessarily, it depends when the frosts start forming. They can form pre-2017 and then sublime during % the 2017-2022 period, and in that case contribute to a temporarily increase in pressure.]
% \bla

%\blu 
Meanwhile, 
%\bla
the 2022 observation confirms the fact that the southern cap has not retreated 
below the 30$^\circ$S latitude level since 1989 from its $\sim$15$^\circ$S extent at that time.
Had it been the case, the surface pressure would have dramatically collapsed since 1989, 
and would be inconsistent with the 2017 and 2022 occultation results.
% \red
% [respond to Emmanuel's comment: ``Add a model case in Fig.~3 to show that?"] -->
% [It is pretty clear from our Fig. 9 when we compare the cases. 
% I don't have a simulation in which the cap retreats during this period.]
% \bla

\begin{acknowledgements}
We thank N. Billot, M. Beck, M. G\"unther, K. Isaak, I. Pagano and the CHEOPS Project Science Office (PSO)
for help in planning observations based on the position of CHEOPS in time.
GBr acknowledges support from CHEOPS ASI-INAF agreement n. 2019-29-HH.0. 
We thank the staff of IAO, Hanle, and CREST, Hosakote that made these observations possible. 
The facilities at IAO and CREST are operated by the Indian Institute of Astrophysics. 
The GROWTH India Telescope (GIT) is a 70-cm telescope with a 0.7-degree field of
view, set up by the Indian Institute of Astrophysics (IIA) and the Indian Institute of Technology
Bombay (IITB) with funding from DST-SERB and IUSSTF. It is located at the Indian Astronomical
Observatory (Hanle), operated by IIA. We acknowledge funding by the IITB alumni batch of 1994, which
partially supports the operations of the telescope.
J.P.N. and D.K.O. acknowledge the support of the Department of Atomic Energy, Government of India, 
under project identification No. RTI 4002.
This study was partly financed by the National Institute of Science and Technology of the e-Universe project (INCT do e-Universo, CNPq grant 465376/2014-2). This study was financed in part by CAPES – Finance Code 001. The authors acknowledge the respective CNPq grants: B.E.M. 150612/2020-6; F.B.R. 314772/2020-0; R.V.M. 307368/2021-1; M.A. 427700/2018-3, 310683/2017-3, 473002/2013-2; J.I.B.C. acknowledges grants 305917/2019-6, 306691/2022-1 (CNPq) and 201.681/2019 (FAPERJ). 
The predictions of this event benefited from unpublished observations made at the Pico dos Dias
Observatory (Brazil), P-LP23.
J.L.O. acknowledges support by spanish project PID2020-112789GBI00 from AEI.
P.S-S. acknowledges financial support from the Spanish I+D+i project PID2022-139555NB-I00 
funded by MCIN/AEI/10.13039/501100011033.
J.L.O. and P.S-S. acknowledge financial support from the Severo Ochoa grant CEX2021-001131-S 
funded by MCIN/AEI/10.13039/501100011033.
R.H.Y. acknowledges the grants from National Natural Science Foundation of China (No. 12073059 \& No.
U2031139) and the National Key R\&D Program of China (No. 2019YFA0405501, 2022YFF0503402). 
Q.Y.Z.  acknowledges the grants from National Natural Science Foundation of China (No. 11988101 \& No.
42075123)
% RSS acknowledges the grants???????????????????????????????????
and the grant from Chinese Academy of Sciences Hundred Talents Project (E12501100C).
\end{acknowledgements}

%    Large tables (longer than one page)
%    Tables larger than a page should be composed at the end of the document.
%    https://www.aanda.org/author-information/latex-issues/latex-examples

\begin{appendix}

\section{Observations}
\label{ap_observations}

\subsection*{{\it CHEOPS Observation}}

% \blu 
%{\bf 
CHEOPS is a dedicated mission for observing exoplanet transits. It is equipped with a 32~cm Ritchey-Chr\'{e}tien telescope with a single, frame-transfer, back-illuminated, $1024 \times 1024$ pixel CCD. There is no filter which results in a bandpass of 0.33-1.13~$\mu$m.
%}
Following the successful CHEOPS observation of a stellar occultation by the trans-Neptunian object Quaoar 
\citep{Mor2022}, predictions were made for various bodies, including Triton. 
Since CHEOPS' predicted orbit is only available at most four months prior to the event date, 
the occultations were predicted statistically, contrarily to ground-based observations. 
%
%{\bf 
CHEOPS is kept in a Sun-synchronous dusk–dawn orbit, 700 km above Earth's
surface\footnote{\url{https://www.esa.int/Science_Exploration/Space_Science/Cheops/Cheops_overview2}}. 
Considering the spacecraft and shadow velocities, the probability of CHEOPS crossing Triton's shadow path was first
estimated to be about 13\%. 
The prediction was continuously updated during the four months before the event 
using new estimations of CHEOPS’ orbits.
Finally, observations were triggered two weeks prior to the event, which  was successfully recorded 
with an exposure time of 3~s.
%}
%Considering CHEOPS is kept in a Sun-synchronous dusk–dawn orbit 700 km above Earth\footnote{\url{https://www.esa.int/Science_Exploration/Space_Science/Cheops/Cheops_overview2}}, we understand that the spacecraft could be located anywhere in this restricted orbit. Then, considering the CHEOPS' and shadow's velocities, we estimated the probability of CHEOPS being under Triton's shadow at some point along its movement to be around 13\%. This proposal was granted time at the third Announcement of Opportunity (AO3) for the CHEOPS Guest Observers Programme\footnote{\url{https://www.cosmos.esa.int/web/cheops-guest-observers-programme/ao-3-programmes}}.

%Within four months of the occultation, when the predicted orbital path of CHEOPS was available, we identified that, indeed, CHEOPS could observe positively the occultation by Triton. The prediction was updated several times as new predicted orbits were available, all of them confirming the conditions for detection. Finally, we triggered the observation two weeks before the occultation, which was positively observed (see \autoref{fig_geometry}).

\begin{figure}[h!]
\centering
\includegraphics[width=\linewidth]{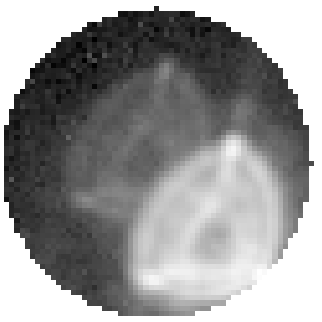}
\hspace{1.0cm}
\caption{%
An imagette from CHEOPS showing the star being occulted by Triton in the centre and 
the bright PSF of Neptune to the lower right. 
The peculiar shape of the PSF is due to the defocussed optics of CHEOPS. 
The diameter of the imagette is 60 arcsecs and the separation between Neptune and Triton 
varied between 18 and 15 arcsecs during the observation.
}%
\label{fig_CHEOPS}
\end{figure}

The defocussed point-spread function (PSF) of CHEOPS results in the nearby Neptune (Fig.~\ref{fig_CHEOPS}) strongly contaminating the standard aperture photometry as derived from the CHEOPS Data Reduction Pipeline \citep[DRP, ][]{hoy22}. To disentangle the photometry and furthermore take advantage of the shorter cadence of the imagettes (3\,s in contrast to the standard 42\,s used for the subarray photometry), we extracted PSF photometry from the imagettes using the Python package PIPE\footnote{\url{https://github.com/alphapsa/PIPE}} 
(PSF Imagette Photometric Extraction, see \citealt{mor21, bra22} for more details). 

%\blu
The resulting imagette photometry clearly resolves the ingress and egress.
% with a photometric noise of \blu 0.7\% \bla per data point at 3~s cadence.
%
The photometric noise is estimated by PIPE to increase from 0.5\% to 0.9\% per exposure
during the 15~min centered on the occultation. 
This noise is mainly due to the photon counting statistics, and is
dominated by scattered light from the nearby Neptune, with negligible contribution from the detector.
The noise varies in time due to the asymmetric PSF of CHEOPS in combination with field rotation changing 
with time, that causes a variable contamination from the planet.
%\bla

\subsection*{{\it IAO Observations}}
%
%Observation was attempted from three Indian facilities. No data could be obtained from the 3.6-m {\it Devasthal Optical Telescope} (DOT) at Devasthal due to bad weather.
%
Successful observations were carried out at the Indian Astronomical Observatory (IAO), 
atop Mt. Saraswati, Digpa Ratsa Ri in Hanle, Ladakh using 
the 2-m {\it Himalayan Chandra Telescope} (HCT) and
the 0.7-m robotic {\it GROWTH-India Telescope} (GIT, \citealt{kum22}). 
HCT recorded the event in the J-band with the TIRSPEC instrument. %(Table~\ref{tab_observations}). 
It has a Teledyne 1024$\times$1024 pixel Hawaii-1 PACE array with four quadrants and 
field of view (FoV) of 307\arcsec$\times$307\arcsec and plate scale of 0.3\arcsec per pixel. 
TIRSPEC offers the flexibility of sub-array acquisition for faster readout \citep{Nina14}. 
For this event, a 307\arcsec$\times$20\arcsec sub-array was used to accommodate the target and the reference star 
in adjacent quadrants. 
The detector was readout non-destructively in the up-the-ramp (UTR) mode. 
The exposure time of each UTR cycle was 32~s, with dead-times of 6.3~s between UTR cycles and integration time between consecutive non-destructive readout of 0.21~s. Dark and flat frames were obtained in the same configuration.
Photometric calibration of the occulted star was carried out under similar observing conditions on 5 October 2022.
However, with GIT, only the egress portion of the event could be captured. Observations were carried out in R-band with the $4096 \times 4108$ pixel Andor iKon-XL 230 CCD camera, which has a FoV of $0.7^{\circ}$. The camera was operated in the fast readout (1.0778~s) mode with exposure time of 3~s. 

%\blu
For each frame of HCT and GIT observations, the photometric error was computed by PRAIA \citep{ass23b} 
using a standard procedure based on the signal-to-noise ratio. 
Then, by calibrating the Triton flux using a close-by star, the flux ratio error was obtained 
from the propagated individual errors as 9\% and 5\% for HCT and GIT, respectively.
%from the propagated individual errors as 8\% and 9\% for HCT and GIT, respectively.
%The noise of both HCT and GIT light curves is dominated by the Earth atmospheric scintillation, with negligible contribution from the detectors. The noise is then estimated from the standard deviation of the flux outside the occultation event.
%\bla

\subsubsection*{{\it Yanqi Lake Observation}}
%{\bf 
F. D. Romanov contacted J.Y. Zhao who sent an observation request to observers at Yanqi Lake Observatory,
pertaining to the University of Chinese Academy of Sciences (UCAS, Beijing, China). 
The event was successfully recorded at this station on behalf of F. D. Romanov, using the 0.7-m f/6.5 Corrected Dall-Kirkham (PlaneWave CDK700) telescope equipped with an Andor iKon-L DZ936-BV CCD. 
The CCD was operated in the R-band, using the sub-sampling and fast readout mode (5~MHz), 
with an exposure time of 1~s and a readout time of 0.8~s.
%\blu
Using the same approach as for the IAO observations, the flux ratio error was estimated to be around 4\%.

\begin{table*}[!h]
    \setlength{\tabcolsep}{2mm}
    \renewcommand{\arraystretch}{1}   % spacing between table lines
    \centering
    \caption{%
    Circumstances of observations.
    \label{tab_observations}
    }%
    \begin{tabular}{lllll}
    \hline \hline
    
    Site    & Coordinates   & Telescope aperture (m)    & Exp. time/Cycle (s) & Observers \\
            & Altitude (m)  & Instrument/filter         & or comments         &           \\
    \hline

    \multicolumn{5}{c}{Positive observations} \\
    \hline
    
    CHEOPS  & See Fig.~\ref{fig_geometry}   & 0.32                          & 3.0/3.024 & A. R. Gomes-Júnior    \\
    Space   &                               & broadband 0.33-1.13 $\mu$m    &            & B. E. Morgado            \\
                    
& & & & \\   
                 
    Mt. Saraswati   & 78 57 49.8 E  &  2.0                             & 0.21/0.21  & B. C. Bhatt       \\
    HCT             & 32 46 46.4 N  &  TIFR Near Infrared Spectrometer &            & S. Pramod Kumar   \\
    India           & 4520          &  and Imager (TIRSPEC)/J          &            &                   \\

& & & & \\   

    Mt. Saraswati   & 78 57 52.6 E  &  0.7                              & 3/4.0778          & V. Swain \\
    GROWTH          & 32 46 45.2 N  &  CCD Andor iKon-XL/R              &                   &           \\
    India           & 4517          &                                   &                   &           \\

& & & & \\   

    Yanqi Lake      & 116 40 14.0 E &  0.7                              & 1/1.8             & F. D. Romanov \\
    China           & 40 24 29.3 N  &  CCD Andor iKon-L                 &                   & Y. M. Mao \\
                    & 96            &  DZ936-BV                        &                   & R. H. Ye \\
                    &               &  Johnson-Cousins Rc               &                   & Q. Y. Zou \\
                    &               &                                   &                   & Y. K. Sun \\
                    &               &                                   &                   & Y. Y. Shen \\
                    &               &                                   &                   & J. Y. Zhao \\
    \hline 
    \multicolumn{5}{c}{Observations with weather problems} \\
    \hline
    Devasthal       & 79 41  03.6 E & 3.6 \& 1.3                        & Clouded out   & A. Tej    \\
    India           & 29 21 39.4 N  & TIRCAM2/H \&                      &               & S. Sharma \\
                    & 2450          & ANDOR DZ436/I                     &               & A. Saha   \\
& & & & \\
    
    Nakhodka        & 132 39 24.2 E &  0.355                            & Clouded out       & F. D. Romanov \\
    Russia          & 42 52 02.6 N  &  CMOS Canon EOS 6D                &                   & D. N. Grishin \\
                    & 2             &                                   &                   & L. V. Romanova \\ 

& & & & \\
Mt. Abu             & 72 46 45.2 E  & 1.2 & Clouded out & J. K. Jain    \\
India               & 24 29 17.3 N  & LISA CCD/white    &           &   \\
                    & 1680          &                   &           &   \\

& & & & \\
Udaipur             & 73 40 26.4 E & 0.5                            & Clouded out   & S. K Mathew  \\
India               & 24 36 15.5 N & Multi-Application              &               &               \\
                    & 610          & Solar Telescope (MAST)/white   &               &               \\
& & & & \\
Dhanari             & 72 55 39.6 E & 0.2 & Clouded out & A. Deshpande \\
India               & 24 40 55.5 N  & QHY5L-ii Mono & & S. Deshmukh \\
                    &   369           &              &  & N. Dhyani\\
                    &              &               &   & A. Gokhale\\
                    &              &               &   & A. Kate \\
& & & & \\
                    
za9pya              & 130 38 53.0 E & 0.11              & Clouded out   & R. Kukita \\
Japan               & 31 52 04.4 N  & eVscope v1.0      &               & \\
                    & 335           &                   &               & \\
& & & & \\
26e7vr              & 139 18 49.7 E & 0.11              & Clouded out   & K. Fukui \\
Japan               & 37 44 16.2 N  & eVscope v1.0      &               & \\
                    & 115           &                   &               & \\
& & & & \\
v8vjs9              & 139 18 58.7 E & 0.11              & Clouded out   & K. Fukui \\
Japan               & 37 44 11.6 N  & eVscope v2.0      &               & \\
                    & 0             &                   &               & \\
    \hline                        
\end{tabular}
\end{table*}

%%%%%%%%%%%%%%%%%%%%%%%%%%%%%%%%%%%%%%%%%%%%%%%%%%%%%%%%%%%%%%%

\end{appendix}

\end{document}